\documentclass[fleqn,twoside]{article}
\usepackage{espcrc2}

\usepackage{epsfig}

\newcommand{\AmS}{{\protect\the\textfont2
  A\kern-.1667em\lower.5ex\hbox{M}\kern-.125emS}}

\title{Resolving the $p^+ = 0$ Ambiguity in a Homogeneous Electric Background}

\author{R. P. Woodard\address{Department of Physics, University of Florida \\
        Gainesville, FL 32611, USA}\thanks{e-mail: woodard\@phys.ufl.edu} }

\begin{document}

\begin{abstract}
I present an exact solution for the Heisenberg picture, Dirac 
electron in the presence of an electric field which depends arbitrarily upon
the light cone time parameter $x^+ = (t+x)/\sqrt{2}$. This is the largest class
of background fields for which the mode functions have ever been obtained. The
solution applies to electrons of any mass and in any spacetime dimension. The
traditional ambiguity at $p^+ = 0$ is explicitly resolved. It turns out that 
the initial value operators include not only $(I + \gamma^0 \gamma^1) \psi$ at
$x^+ = 0$ but also $(I - \gamma^0 \gamma^1) \psi$ at $x^- = -L$. Pair creation 
is a discrete and instantaneous event on the light cone, so one can compute the
particle production rate in real time. In $D=1+1$ dimensions one can also see
the anomaly. Another novel feature of the solution is that the expectation 
value of the current operators depends nonanalytically upon the background
field. This seems to suggest a new, strong phase of QED.
\vspace{1pc}
\end{abstract}

\maketitle

\section{Introduction}

I will be reporting on work done with my good friends Nikolaos Tsamis and 
Theodore Tomaras, both from the University of Crete. What we did is to 
solve the Dirac equation for the electron field operator in the presence 
of an electric background field $\vec{E} = E(x^+) \widehat{x}$ which 
depends arbitrarily upon the light cone time parameter $x^+$ \cite{TTW1,TTW2}.
As I have reported elsewhere \cite{RPW}, this system provides a wonderfully
explicit example of particle production, the first ever check of the axial
vector anomaly for the massive theory in $D=1+1$ dimensions, and the hint of 
an infinite volume phase transition in the fact that expectation values of 
the current operators depend nonanalytically upon the background field. Here
I discuss what it tells us about resolving the $p^+ = 0$ ambiguity in 
light cone quantum field theory.

We define the light cone coordinates as $x^{\pm} \equiv (t \pm x)/\sqrt{2}$.
The remaining, ($D-2 \equiv d$) transverse coordinates are denoted thusly: 
$x_{\bot}$. We work in the gauge where $A_+ = 0$ and,
\begin{equation}
A_-(x^+) = - \int_0^{x^+} du E(u) \; . \label{Aminus}
\end{equation}
The transverse components of the vector potential vanish, $A_{\bot} = 0$.
It is useful as well to define $\pm$ spinor components,
\begin{equation}
\psi_{\pm}(x^+,x^-,x_{\bot}) \equiv P_{\pm} \psi(x^+,x^-,x_{\bot}) \; ,
\end{equation}
where $P_{\pm} \equiv (I \pm \gamma^0 \gamma^1)/2$. With these conventions 
the Dirac equation takes the form,
\begin{eqnarray}
i \partial_+ \psi_+ & = & \frac12 \left(m + i \nabla_{\bot} \cdot 
\gamma_{\bot}\right) \gamma^- \psi_- , \\
(i \partial_- - e A_-) \psi_- & = & \frac12 \left(m + i \nabla_{\bot} \cdot 
\gamma_{\bot}\right) \gamma^+ \psi_+ .
\end{eqnarray}

\section{The physics of the ambiguity}

Let us denote the Fourier transform on the transverse coordinates by a tilde,
\begin{eqnarray}
\lefteqn{\widetilde{\psi}_{\pm}(x^+,x^-,k_{\bot}) \equiv } \nonumber \\
& & \int d^dx_{\bot} e^{-i k_{\bot} \cdot x_{\bot}} \psi_{\pm}(x^+,
x^-,x_{\bot}) \; .
\end{eqnarray}
A hat denotes additionally transforming on $x^-$,
\begin{eqnarray}
\lefteqn{\widehat{\psi}_{\pm}(x^+,k^+,k_{\bot}) \equiv } \nonumber \\
& & \int_{-\infty}^{\infty} dx^- e^{i k^+ x^-} \widetilde{\psi}_{\pm}(x^+,
x^-,k_{\bot}) \; .
\end{eqnarray}
The Fourier transformed field equations are,
\begin{eqnarray}
i \partial_+ \widehat{\psi}_+ & = & \frac12 \left(m - k_{\bot} \cdot \gamma_{
\bot}\right) \gamma^- \widehat{\psi}_- , \label{psihat+} \\
(k^+ - e A_-) \widehat{\psi}_- & = & \frac12 \left(m - i k_{\bot} \cdot 
\gamma_{\bot}\right) \gamma^+ \widehat{\psi}_+ . \label{psihat-}
\end{eqnarray}

Multiplying (\ref{psihat+}) by $[k^+ - e A_-(x^+)]$ and then using 
(\ref{psihat-}) gives the following equation for $\widehat{\psi}_+$,
\begin{equation}
(k^+ - e A_-) i \partial_+ \widehat{\psi}_+ = \frac12 \omega^2_{\bot} 
\widehat{\psi}_+ \; , 
\end{equation}
where $\omega^2_{\bot} \equiv k_{\bot} \cdot k_{\bot} + m^2$. The solution is,
\begin{eqnarray}
\lefteqn{\widehat{\psi}_+(x^+,k^+,k_{\bot}) = \widehat{\psi}_+(0,k^+,k_{\bot})
} \nonumber \\
& & \times \exp\left[-\frac{i}2 \omega^2_{\bot} \int_0^{x^+} {du \over k^+ - 
e A_-(u)} \right] \; . \label{brandX}
\end{eqnarray}
This is the traditional solution form in light cone field theory, with the 
field evolved forward from initial value data specified on the surface at 
$x^+ = 0$. Contact with familiar physics can be made by setting $A_-$ to zero,
\begin{equation}
\exp\left[\int_0^{x^+} {-i \omega^2_{\bot} du/2 \over k^+ - e A_-(u)}\right]
\longrightarrow e^{- i k^- x^+} \;, \label{familiar}
\end{equation}
where $k^- = \omega^2_{\bot}/{2 k^+}$.

Suppose now that the electric field is positive. Since the electron charge
is negative, it follows from (\ref{Aminus}) that the vector potential is
an increasing function of $x^+$. For some values of $k^+ > 0$ (all of them if
$e A_-(x^+)$ increases without bound) there comes a value of the light cone 
time parameter at which $k^+ - e A_-(x^+) = 0$. The solution (\ref{brandX}) 
does not exist beyond this point. Note from (\ref{familiar}) that this is the 
same problem which occurs at $k^+ = 0$ in the case of zero background. What 
the electric field does is to prevent the problem from remaining localized at
a single value of $k^+$.

The reason a problem {\it can} arise is because we should not have Fourier 
transformed with respect to $x^-$. For a Fourier transform to exist, the 
function being transformed must fall off at large values. This poses no 
problem for the transverse coordinates. If $\psi_{\pm}$ falls off for large 
$\Vert x_{\bot} \Vert$ at $x^+ =0$ then causality guarantees that it will 
continue to fall off for all $x^+ > 0$. So if $\widetilde{\psi}_{\pm}(0,x^-,
k_{\bot})$ exists then we can expect $\widetilde{\psi}_{\pm}(x^+,x^-,k_{\bot})$ 
to exist as well. But the $x^-$ axis is lightlike; causality poses no barrier 
to propagation in the positive $x^-$ direction over an arbitrarily short 
period of $x^+$. Hence the existence of $\widehat{\psi}_{\pm}(0,k^+,k_{\bot})$ 
in no way guarantees the existence of $\widehat{\psi}_{\pm}(x^+,k^+,k_{\bot})$.

The reason a problem {\it does} arise is that the homogeneous electric 
background induces the production of $e^+ e^-$ pairs. Because the electric
field points along the positive $x$ axis, the electron is accelerated in
the negative $x$ direction. As it approaches the speed of light, its motion
becomes parallel to the $x^-$ axis, {\it and it leaves the manifold}. The
process is depicted in Fig.~1. 

\centerline{\psfig{figure=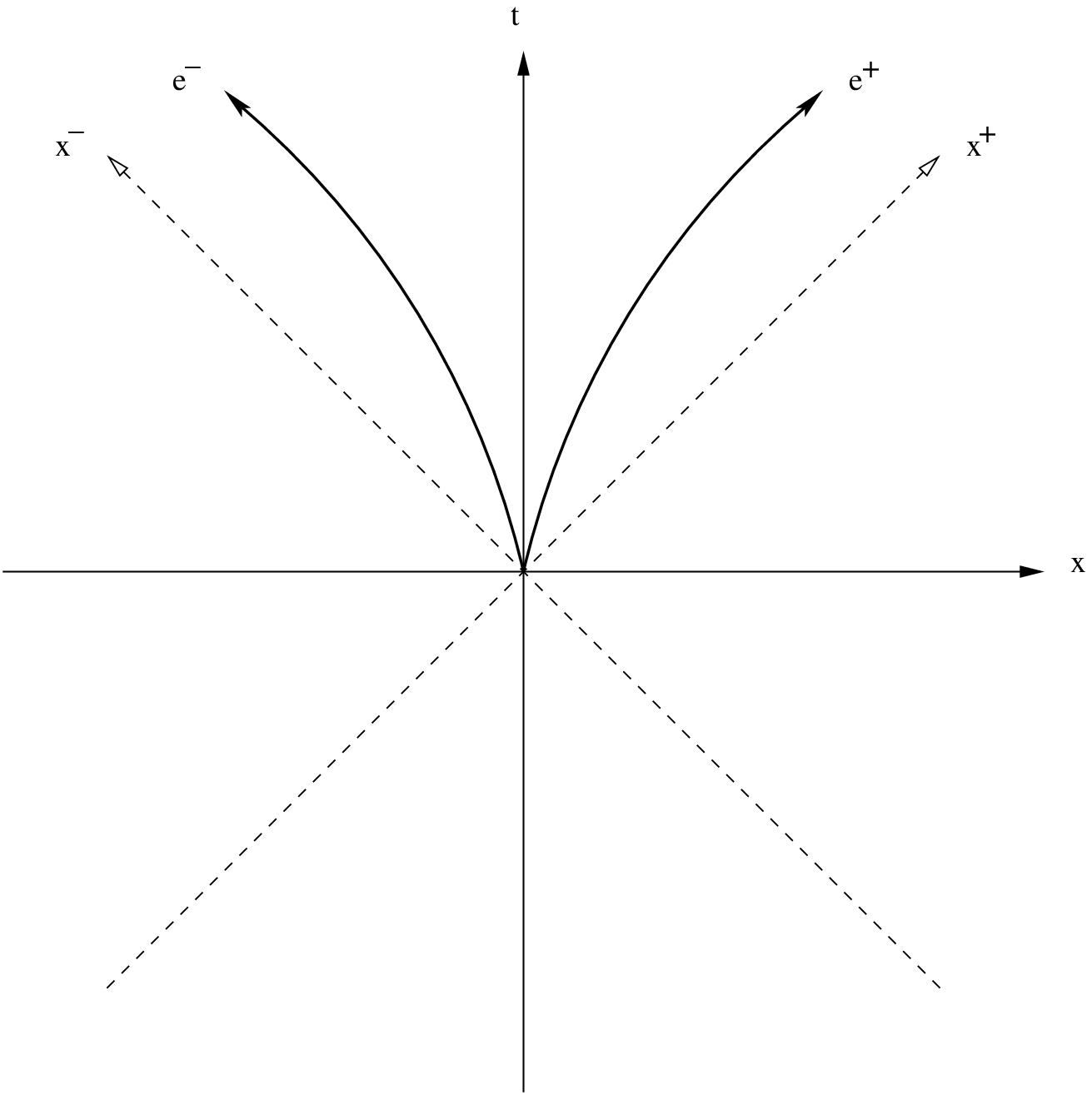,height=7.5cm}}

\centerline{Fig.~1: The evolution of a virtual $e^+ e^-$ pair.}

It turns out that electrons are {\it created} moving at the speed of light,
and that the instant of creation for mode $k^+$ occurs at precisely $k^+ = 
e A_-(x^+)$. Why this is so can be understood from the fact that evolving form 
a prepared state at $x^+ = 0$ is the infinite boost limit of evolving from a 
prepared state at $t' = 0$ \cite{Kogut}. Suppose we consider a primed frame 
in which the empty state is specified at $t'=0$. The homogeneous electric 
field will result in particle creation, but the created particles will possess 
finite, nonzero $p^{\prime \pm}$. Now consider the light cone momenta in the 
frame obtained by boosting to velocity $\beta$ in the $x$ direction,
\begin{equation}
p^+ = \sqrt{{1 - \beta \over 1 + \beta}} p^{\prime +} \quad , \quad
p^- = \sqrt{{1 + \beta \over 1 - \beta}} p^{\prime -} \; .
\end{equation}
As $\beta$ approaches unity we see that $p^+$ goes to zero. Therefore any
particle created in the light cone problem comes out with $p^+ = 0$. But the
physical momentum in a background vector potential is the minimally coupled
one, $p^+ = k^+ - e A_-(x^+)$. Hence pair creation occurs at the instant when
$k^+ = e A_-(x^+)$.

\section{The unambiguous solution}

The correct procedure is to solve for $\widetilde{\psi}_{\pm}$ on the domain 
bounded by $x^+ > 0$ and $x^- > -L$, as shown in Fig.~2. The field equations
are,
\begin{eqnarray}
i \partial_+ \widetilde{\psi}_+ & = & \frac12 \left(m - k_{\bot} \cdot \gamma_{
\bot}\right) \gamma^- \widetilde{\psi}_- , \label{psitilde+} \\
(i \partial_- - e A_-) \widetilde{\psi}_- & = & \frac12 \left(m - k_{\bot} 
\cdot \gamma_{\bot}\right) \gamma^+ \widetilde{\psi}_+ . \label{psitilde-}
\end{eqnarray}
By integrating (\ref{psitilde+}) with respect to $x^+$  and (\ref{psitilde-})
with respect to $x^-$ we obtain,
\begin{eqnarray}
\lefteqn{\widetilde{\psi}_+(x^+,x^-,k_{\bot}) = \widetilde{\psi}_+(0,x^-,
k_{\bot}) - \frac{i}2 (m } \nonumber \\
& & - k_{\bot} \cdot \gamma_{\bot}) \gamma^- \int_0^{x^+} du 
\widetilde{\psi}_-(u,x^-,k_{\bot}) \label{int1} \; , \\
\lefteqn{\widetilde{\psi}_-(x^+,x^-,k_{\bot}) = } \nonumber \\
& & e^{-ieA_-(x^+) (x^- + L)} \widetilde{\psi}_-(x^+,-L,k_{\bot}) \nonumber \\
& & - \frac{i}2 \left(m - k_{\bot} \cdot \gamma_{\bot}\right) \gamma^+
\int_{-L}^{x^-} dv \nonumber \\ 
& & \quad \times e^{-ieA_-(x^+) (x^- - v)} \widetilde{\psi}_+(x^+,v,k_{\bot}) 
\; . \label{int2}
\end{eqnarray}
Substituting (\ref{int2}) into (\ref{int1}) and iterating, one is led to an
infinite series solution for $\widetilde{\psi}_+$ which can be summed. After 
some work one obtains \cite{TTW2},
\begin{eqnarray}
\lefteqn{\widetilde{\psi}_+(x^+,x^-,k_{\bot}) = } \nonumber \\
& & \int_{-L}^{\infty} dy^- \int_{-\infty}^{+\infty} {dk^+ \over 2 \pi} 
e^{i (k^+ + i/L) (y^- - x^-)} \nonumber \\
& & \times \Bigl\{{\cal E}(0,x^+;k^+,k_{\bot}) \widetilde{\psi}_+(0,y^-,
k_{\bot}) -\frac{i}2 (m \nonumber \\
& & - k_{\bot} \cdot \gamma_{\bot}) \gamma^- \int_0^{x^+} dy^+ e^{-ieA_-(y^+) 
(y^- + L)} \nonumber \\
& & \times {\cal E}(y^+,x^+;k^+,k_{\bot}) \widetilde{\psi}_-(y^+,-L,k_{\bot}) 
\Bigr\} \; , \label{psi+}
\end{eqnarray}
where the $E$-dependent mode function is,
\begin{eqnarray}
\lefteqn{{\cal E}[eA_-](y^+,x^+;k^+,k_{\bot}) \equiv } \nonumber \\
& & \exp\left[- {i\over 2} \omega^2_{\bot} \int_{y^+}^{x^+} {du \over k^+ - 
eA_-(u) + i/L} \right] \; .
\end{eqnarray}
The field $\widetilde{\psi}_-$ is obtained by differentiating 
$\widetilde{\psi}_+$,
\begin{eqnarray}
\lefteqn{\widetilde{\psi}_-(x^+,x^-,k_{\bot}) =} \nonumber \\
& & \left({m - k_{\bot} \cdot \gamma_{\bot} \over \omega^2_{\bot}} \right) 
\gamma^+ i \partial_+ \widetilde{\psi}_+(x^+,x^-,k_{\bot}) \; . \label{psi-}
\end{eqnarray}

\centerline{\psfig{figure=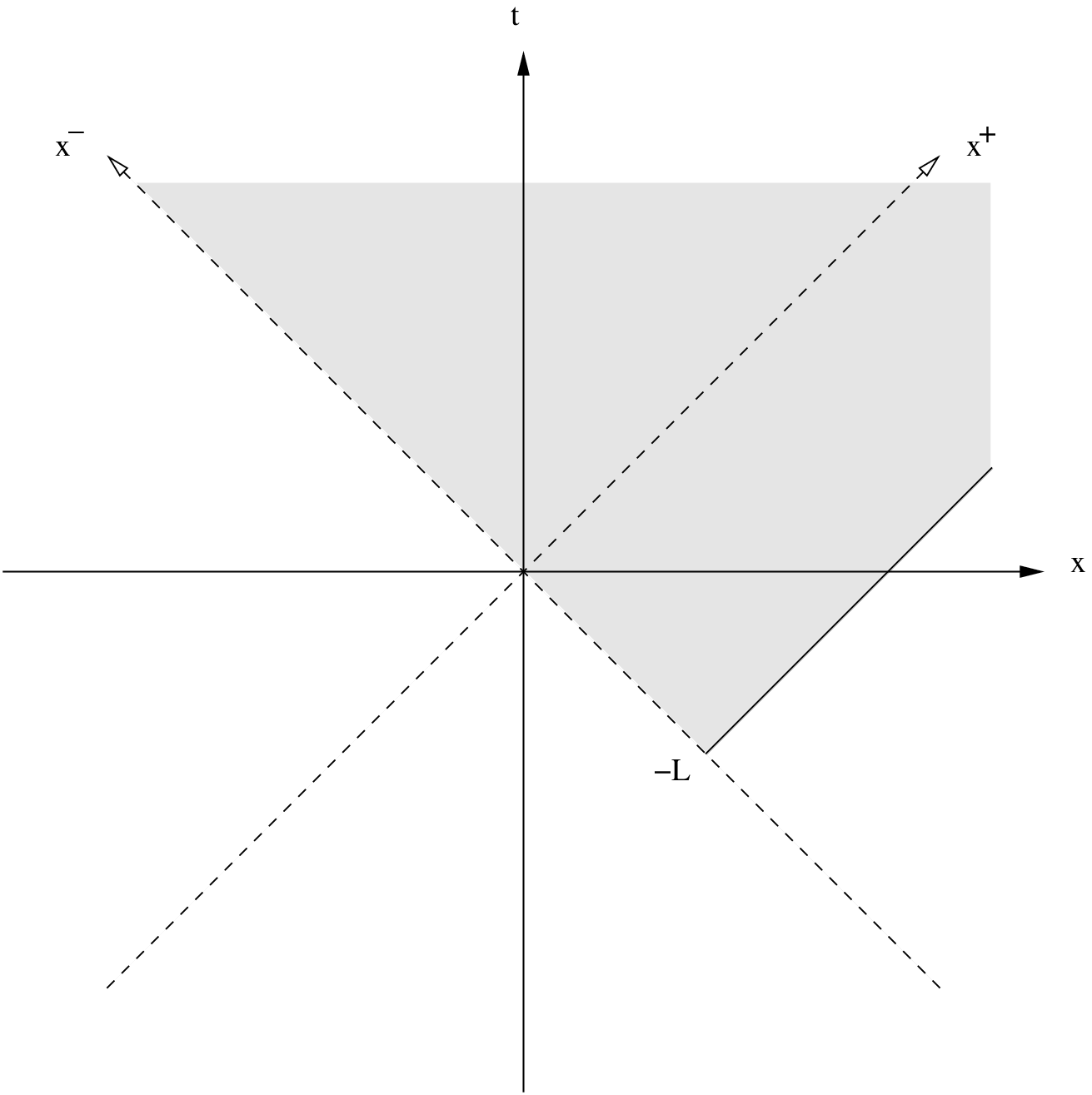,height=7.5cm}}

\centerline{Fig.~2: The domain of solution is shaded.}

Even though the Fourier transform with respect to $x^-$ does not exist,
(\ref{psi+}) is something very like it. However, the ambiguities of the
previous section have been resolved through three key replacements. The
first of these precisely defines the singularity at $k^+ = e A_-(x^+)$,
\begin{equation}
k^+ \longrightarrow k^+ + \frac{i}{L} \; .
\end{equation}
The second replacement can be regarded as defining a modified Fourier
transform which is cut off in the negative $x^-$ direction and whose 
convergence at large $x^-$ is enforced by factors of $i/L$,
\begin{eqnarray}
\lefteqn{\int_{-\infty}^{\infty} {dk^+ \over 2 \pi} e^{-i k^+ x^-} 
\int_{-\infty}^{\infty} dy^- e^{i k^+ y^-} \longrightarrow } \nonumber \\
& & \int_{-L}^{\infty} dy^- \int_{-\infty}^{\infty} {dk^+ \over 2\pi} 
e^{i (k^+ + i/L) (y^- - x^-)} \; .
\end{eqnarray}
The third replacement is profound because it alters the initial value operators
which are the true degrees of freedom of any quantum field theory. In addition 
to providing $\widetilde{\psi}_+(0,x^-,k_{\bot})$ for $x^- > -L$ we must also 
give $\widetilde{\psi}_-(x^+,-L,k_{\bot})$ for $x^+ > 0$. 

Canonical quantization shows the nonzero part of the anti-commutation algebra 
to be \cite{TTW1},
\begin{eqnarray}
\lefteqn{\left\{\psi_+(0,x^-,x_{\bot}),\psi_+^{\dagger}(0,y^-,y_{\bot})\right\}
= } \nonumber \\
& & \frac{P_+}{\sqrt{2}} \delta(x^- - y^-) \delta^d(x_{\bot} - y_{\bot}) 
\; , \label{CA++} \\
\lefteqn{\left\{\psi_-(x^+,-L,x_{\bot}),\psi_-^{\dagger}(y^+,-L,y_{\bot})
\right\} = } \nonumber \\
& & \frac{P_-}{\sqrt{2}} \delta(x^- - y^-) \delta^d(x_{\bot} - y_{\bot}) 
\; . \label{CA--}
\end{eqnarray}
We assume that the initial value operators act upon ``the vacuum'' in the same 
way as they do for zero electric field. Computing the VEV of any operator 
therefore consists of first employing (\ref{psi+}-\ref{psi-}) to express that 
operator in terms of the initial value operators, and then taking the 
expectation value of these in the free theory. It is often useful to take 
the large $L$ limit as well.

\section{$\widetilde{\psi}_-(x^+,-L,k_{\bot})$ matters!}

The need for independent operators on a surface of constant $x^-$ is obvious 
for the massless case in $D = 1 + 1$ dimensions \cite{NR,GMC1} because 
otherwise the left-movers would be missing. Although the necessity for
including these operators for general $m$ and $D$ has been argued \cite{GMC2} 
there is a widespread belief that this can be avoided by imposing boundary 
conditions. That is true for the free theory in the trivial background 
\cite{HW} because then the problem resides at $p^+ = 0$ and there is never
any mixing of modes. Our electric background spoils this by causing the
problem at $p^+ = 0$ to migrate ever higher up along the $k^+$ axis. The 
effect can be quantified by comparing important relations computed with 
$\widetilde{\psi}_-(x^+,-L,k_{\bot})$ fully dynamical and again with it 
constrained to zero.

Let us start with the canonical anti-com\-mu\-tator between $\widetilde{\psi
}_+$ and $\widetilde{\psi}^{\dagger}_+$ at fixed $x^+$. Each of the operators 
consists of a term from the surface at $x^+ = 0$ and another from the surface
at $x^- = -L$. After using (\ref{CA++}) and (\ref{CA--}) and performing the
dummy integrations over $y^-$ and $y^{-\prime}$ we obtain,
\begin{eqnarray}
\lefteqn{\left\{ \widetilde{\psi}_+(x^+,x^-,k_{\bot}),\widetilde{\psi}_+^{
\dagger}(x^+,x^{-\prime},k_{\bot}^{\prime})\right\} = {P_+ \over \sqrt{2}} 
(2\pi)^d }\nonumber \\
& & \times \delta^d(k_{\bot} - k_{\bot}^{\prime}) \int_{-\infty}^{\infty} 
{dk^+ \over 2 \pi} e^{-i (k^+ + i/L) (x^- + L)} \nonumber \\
& & \times \int_{-\infty}^{\infty} {dk^{+\prime} \over 2 \pi} e^{-i (k^{+
\prime} + i/L) (x^{-\prime} + L)} \nonumber \\
& & \times \left\{ {i {\cal E}\left(0,x^+;k^+,k_{\bot}\right) {\cal E}^*\left(
0,x^+;k^{+\prime},k_{\bot}\right) \over k^+ - k^{+\prime} + 2i/L} \right.
\nonumber \\
& & \quad - {\omega^2_{\bot} \over 2} \int_0^{x^+} dy^+ {i {\cal E}\left(
y^+,x^+;k^+;k_{\bot}\right) \over k^+ - e A_-(y^+) + i/L} \nonumber \\
& & \hspace{2cm} \left. \times {i {\cal E}^*\left(y^+,x^+;k^{+\prime};k_{
\bot}\right) \over k^{+\prime} - e A_-(y^+) - i/L} \right\} . \label{com}
\end{eqnarray}
The $y^+$ integrand is a total derivative,
\begin{eqnarray}
\lefteqn{- {\omega^2_{\bot} \over 2} {i {\cal E}\left(y^+,x^+;k^+;k_{\bot}
\right) \over k^+ - e A_- + i/L} {i {\cal E}^*\left(y^+,x^+;k^{+\prime};k_{
\bot}\right) \over k^{+\prime} - e A_- - i/L} = } \nonumber \\
& & {\partial \over \partial y^+} \left[ {i {\cal E}(y^+,x^+;\dots) 
{\cal E}^*(y^+,x^+;\dots) \over k^+ - k^{+\prime} + 2i/L}
\right] .
\end{eqnarray}
The lower limit cancels the first ($++$) term in (\ref{com}) leaving only the
upper limit at which the mode functions become unity,
\begin{eqnarray}
\lefteqn{\left\{ \widetilde{\psi}_+(x^+,x^-,k_{\bot}),\widetilde{\psi}_+^{
\dagger}(x^+,x^{-\prime},k_{\bot}^{\prime})\right\} = {P_+ \over \sqrt{2}} 
(2\pi)^d }\nonumber \\
& & \times \delta^d(k_{\bot} - k_{\bot}^{\prime}) \int_{-\infty}^{\infty} 
{dk^+ \over 2 \pi} e^{-i (k^+ + i/L) (x^- + L)} \nonumber \\
& & \times \int_{-\infty}^{\infty} {dk^{+\prime} \over 2 \pi} {e^{-i (k^{+
\prime} + i/L) (x^{-\prime} + L)} \over k^+ - k^{+ \prime} + 2i/L} \; , \\
& & = {P_+ \over \sqrt{2}} \delta(x^- - x^{-\prime}) (2\pi)^d \delta^d(
k_{\bot} - k_{\bot}^{\prime}) \; .
\end{eqnarray}

Had the operators at $x^- = -L$ not been dynamical there would be no second 
($--$) term in (\ref{com}). In that case the mode functions do not drop
out and one experiences a loss of amplitude as the exponent picks up a 
negative real part by integrating through $k^+ = e A_-(u)$. In the large 
$L$ limit the result is,
\begin{eqnarray}
\lefteqn{{P_+ \over \sqrt{2}} (2\pi)^d \delta^d(k_{\bot} - k_{\bot}^{\prime}) 
\Biggl\{\delta(x^- - x^{-\prime}) - \int_0^{eA_-(x^+)} } \nonumber \\
& & \times {dk^+ \over 2\pi} e^{-i k^+ (x^- - x^{-\prime})} \left[1 - e^{-2 
\pi \lambda(k^+,k_{\bot})}\right] \Biggr\} .
\end{eqnarray}
The function $\lambda(k^+,k_{\bot})$ is,
\begin{equation}
\lambda(k^+,k_{\bot}) \equiv {\omega^2_{\bot} \over 2 \vert e E(X(k^+)) \vert}
\; , 
\end{equation}
and $X(k^+)$ is the light cone time at which $k^+ = e A_-(X)$. We therefore
see that constraining the operators at $x^- = -L$ leads to a progressive loss
of unitarity, even in the limit of infinite $L$. 

Unitarity is not the only thing that goes wrong: one also loses particle
creation and current conservation. Here I will only quote results to save 
space, and I will always take the infinite $L$ limit. With the operators 
at $x^- = -L$ included, the probability of an initially empty state to 
contain a positron of fixed spin, $k^+$ and $k_{\bot}$ is,
\begin{eqnarray}
\lefteqn{{\rm Prob}(x^+,k^+,k_{\bot}) = } \nonumber \\
& & \theta\left(k^+ - e A_-(x^+)\right) e^{-2\pi \lambda(k^+,k_{\bot})} \; . 
\label{prob}
\end{eqnarray}
Without the $x^- = -L$ operators one gets zero! 

With the operators at $x^- = -L$ included the VEV of the current density is,
\begin{eqnarray}
\lefteqn{\Bigl\langle \Omega \Bigl\vert J^+(x^+,x^-,x_{\bot}) \Bigr\vert 
\Omega \Bigr\rangle_{\rm with} = -2^{[\frac{D}2]} e } \nonumber \\
& & \times \int_0^{eA_-(x^+)} {dp^+ \over 2 \pi} \int {d^d p_{\bot} \over 
(2\pi)^d} e^{-2\pi \lambda(p^+,p_{\bot})} \; . \label{J+}
\end{eqnarray}
Without the $x^- = -L$ operators one gets,
\begin{eqnarray}
\lefteqn{\Bigl\langle \Omega \Bigl\vert J^+(x^+,x^-,x_{\bot}) \Bigr\vert 
\Omega \Bigr\rangle_{\rm wout} = - 2^{[\frac{D}2] -1} e } \nonumber \\
& & \times \int_0^{eA_-(x^+)} {dp^+ \over 2 \pi} \int {d^d p_{\bot} \over 
(2\pi)^d} \left[1 + e^{-2\pi \lambda} \right] \; . \label{J+wo}
\end{eqnarray}
The $p_{\bot}$ integral of the first term in the square brackets is not 
even finite for $d > 0$!

The VEV of $J^-$ diverges for infinite $L$ on account of the flux of electrons
leaving the manifold after being created with uniform probability all the way
back along the $x^-$ axis. However, the derivative with respect to $x^-$ is
finite. With the operators at $x^- = -L$ included the result is,
\begin{eqnarray}
\lefteqn{\partial_- \Bigl\langle \Omega \Bigl\vert J^+(x^+,x^-,x_{\bot}) 
\Bigr\vert \Omega \Bigr\rangle_{\rm with} = 2^{[\frac{D}2]} e } \nonumber \\
& & \times {e A_-^{\prime}(x^+) \over 2 \pi} \int {d^d p_{\bot} \over 
(2\pi)^d} e^{-2\pi \lambda(eA_-,p_{\bot})} \; . \label{J-}
\end{eqnarray}
Note that (\ref{J+}) and (\ref{J-}) are consistent with current conservation.
However, constraining the operators at $x^- = -L$ would give,
\begin{eqnarray}
\lefteqn{\partial_- \Bigl\langle \Omega \Bigl\vert J^+(x^+,x^-,x_{\bot}) 
\Bigr\vert \Omega \Bigr\rangle_{\rm wout} = -2^{[\frac{D}2]-1} e } \nonumber \\
& & \times {e A_-^{\prime}(x^+) \over 2 \pi} \int {d^d p_{\bot} \over 
(2\pi)^d} \left[1 - e^{-2\pi \lambda(eA_-,p_{\bot})} \right] \; . \label{J-wo}
\end{eqnarray}
Discarding the operators at $x^- = -L$ therefore results in a violation of
current conservation!

The existence of these catastrophes when the theory is still free but the 
background is nontrivial indicates that one should not suppress the 
operators at $x^- = -L$ in the interacting theory, even in the limit 
of infinite $L$. This may be a disaster for what we seek to do on the 
light cone. On the other hand, it may simply betoken the need for extra
interactions which result from integrating out the operators at $x^- = -L$.

\centerline{\bf Acknowledgments}

I have profited from discussions with D. S.  Hwang, G. McCartor, M. Soussa, 
T. N. Tomaras and N. C. Tsamis. This work was partially supported by DOE 
contract DE-FG02-97ER\-41029 and by the Institute for Fundamental Theory.

\end{document}